\documentclass[conference]{IEEEtran}
\IEEEoverridecommandlockouts
\usepackage{comment}
\usepackage{cite}
\usepackage{amsmath,amssymb,amsfonts}
\usepackage{algorithmic}
\usepackage{graphicx}
\usepackage{textcomp}
\usepackage{xcolor}
\usepackage{float}
\def\BibTeX{{\rm B\kern-.05em{\sc i\kern-.025em b}\kern-.08em
    T\kern-.1667em\lower.7ex\hbox{E}\kern-.125emX}}
\begin{document}

\title{An Innovative Security Strategy
using Reactive
Web Application Honeypot\\
}
\author{\IEEEauthorblockN{1\textsuperscript{st} Rajat Gupta}
\IEEEauthorblockA{\textit{School of Computer Science} \\
\textit{Vellore Institute of Technology}\\
Vellore, India \\
rajat.gupta2017@vitstudent.ac.in}
\and
\IEEEauthorblockN{2\textsuperscript{nd} Madhu Viswanatham V.}
\IEEEauthorblockA{\textit{School of Computer Science} \\
\textit{Vellore Institute of Technology}\\
Vellore, India \\
vmadhuviswanatham@vit.ac.in}
\and
\IEEEauthorblockN{3\textsuperscript{rd} Manikandan K.}
\IEEEauthorblockA{\textit{School of Computer Science} \\
\textit{Vellore Institute of Technology}\\
Vellore, India \\
kmanikandan@vit.ac.in}

}

\maketitle

\begin{abstract}
Nowadays, web applications have become most prevalent in the industry, and the critical data of most organizations stored using web apps. Hence, web applications a much bigger target for diverse cyber-attacks, which varies from database injections-SQL injection, PHP object injection, template injection, XML external entity injection, unsanitized input attacks- Cross-Site Scripting(XSS) and many more. As mitigation for them, among many proposed solutions, web application honeypots are a much sophisticated and powerful protection mechanism.

In this paper, we propose a low interaction, adaptive, and dynamic web application honeypot that imitates the vulnerabilities through HTTP events. The honeypot is built with SNARE and TANNER; SNARE creates the attack surface and sends the requests to TANNER, which evaluates them and decides how SNARE should respond to the requests. TANNER is an analysis and classification tool, which analyzes and evaluates HTTP requests served by SNARE, and to compose the response, it is powered by emulators, which are engines used for the emulation of vulnerabilities.
\end{abstract}

\begin{IEEEkeywords}
\textit{SQL, XML, Out of Band injection, XXE injection, XSS, asynchronous docker}
\end{IEEEkeywords}

\section{\textbf{Introduction}}
Web Applications are more prone to cyber-attacks, and from the various mechanisms proposed to enhance the security of the applications, the most reliable solution is to use Honeypots. The chief designated purpose of a honeypot is to deceive an attacker by imitating the real application and redirecting the events to different engines to detect possible attacks-types, possible owners, IP addresses, and logs the intrusion information. It is helpful to circumvent the attacker that they have managed to gain access to a real system with real attack scenarios. 
\medskip

Modern protections such as intrusion prevention systems, proxy servers, firewalls are vital pillars in information security. Among them, honeypots are the most sophisticated tool which gets disguised, pretends to be an authentic system, and emulates the vulnerabilities to detect malicious attackers. The idea behind web application honeypot is to create an attack surface containing indexed web pages to attract the attackers, an analysis tool working as a brain of the whole system to analyze all the HTTP events and recognizes the types of attacks. Honeypots are usually deployed in the production network and logs all the malicious activities for the administrator in real-time.

\section{\textbf{Background}}

%\subsection{Maintaining the Integrity of the Specifications}

The major widely used honeypot systems ever built in the information security community are Google Hack Honeypot, Glastopf, and HIHAT. 
The classification of honeypots is based on the level of interaction with the system(low, medium, high). Low interaction honeypots are not vulnerable and cannot be exploited by the attack carried on the vulnerable emulated service. Medium interaction honeypots emulate vulnerable services less complicated than high interaction honeypots but still more than low interaction. Whereas High interaction honeypots emulate real services and vulnerabilities that can be directly accessed by the attackers, and hence these are the most sophisticated in terms of detecting malicious activity on the network. Although high interaction honeypots are inclined to risks, and attackers can compromise the real operating system to own the production environment. \cite{b1}
\medskip

HIHAT-A High Interaction Honeypot Analysis Toolkit depends on existing PHP web applications and transforms them into real honeypots in an automated mode. But this requires a dedicated server, and if an attacker owns the server, then HIHAT cannot protect the production systems from getting attacked. \cite{b5}
\medskip

GHH-Google Hack Honeypot uses the power of Google Dorking to attract attackers by correctly placing invisible links on the website. Attackers try to find this invisible data, which might be fake user's accounts, ports, and many more. Appropriately configured honeypot does not give access to the real backend service and logs the malicious activity. This information is useful to detect the attack pattern and statistics, but this honeypot does not provide real security to the application.\cite{b9}
\medskip

One of the most recent low interaction dynamic web application honeypot is Glastopf \cite{b2}. It emulates the vulnerabilities in web apps, scans the malicious paths (URLs), detects the type of attack. Glastopf has vulnerable web pages that are published on search engines and crawlers to attract malicious activities. Attackers find the indexed path and try to attack the system, which is then detected and gets logged by the honeypot. There are certain limitations of this honeypot-firstly it's the client-side interface which is quite primary, and attackers can easily recognize that this is not a real system. Secondly, it only supports SQL injection, remote (RFI), and local file inclusion(LFI) vulnerabilities.  
\medskip

All of the above honeypots apart from Glastopf use prebuilt web page templates modified from real systems to entice the attackers, but this makes the honeypot system more static and adding new templates is a very time consuming process. Glastopf is more dynamic and adaptive to different environments. But it has basic web pages which makes it hard to camouflage the honeypot.  

\section{\textbf{Proposed Methodology}}
This paper proposes a solution is an advanced dynamic honeypot for web applications, which is an enhanced form of Glastopf. Our solution enhances camouflage capability by generating dynamic real web page templates. It comprises powerful emulators for complex vulnerabilities like XML External Entity Injection (XXE) with OOB (Out of Band injection) support, Template Injection, which supports multiple popular templating engines and many more. It first checks if the HTTP request if it matches the supported vulnerabilities or not, then it executes the query in a safe environment (we are using custom Docker environment) and returns the response to the attacker precisely, almost similar to the real system. It uses vulnerable code templates to create emulators for vulnerabilities; the significant benefit of using this method is that they can tempt attackers efficiently as they are very close to real systems. But this makes the honeypot more static, and to add support for new vulnerabilities, we have to add new emulators, which are a bit tedious manually.
\medskip

The proposed method has various components stitched together to work as an adaptive honeypot-

\begin{figure}[htbp]
\centerline{\includegraphics{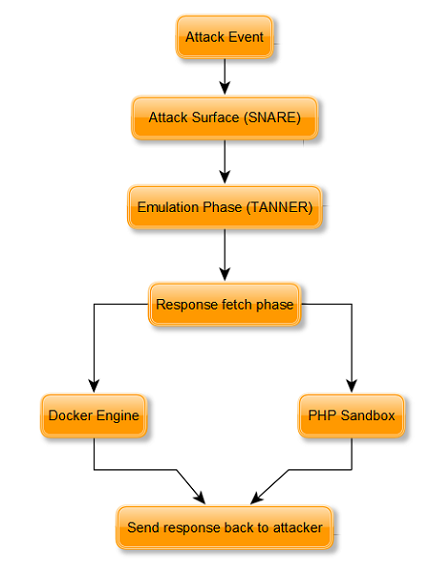}}
\caption{\textit{General functionality overview.}}
\label{fig1}
\end{figure}

\subsection{\textbf{Super Next-Generation Advanced Reactive Honeypot - SNARE}}\label{AA}
The first and principal thing we need is the attack surface with all the emulated vulnerabilities. Here, we introduce SNARE; it creates the attack surface by cloning all the web pages of the application fed as input and names the files by their md5 hash. It serves all the web pages on top of itself, becoming a server and hence monitoring all the HTTP events/flows throughout the application. SNARE has an inbuilt Cloner that works asynchronously to reduce the time taken to clone all the pages but has to be invoked before SNARE to serve all the pages. Cloner is also capable of handling relative URLs as well. Suppose if the requests are localhost/foo, localhost/foo/bar, and suppose the root is localhost it first clones /foo page and when it encounters /foo/bar page it clones /bar after setting localhost/foo as root. Moreover, it also scrapes all the images on the web pages, scripts, and action elements as well so that the clone looks as good as the real system. If an attacker fingerprints SNARE, it deceives them and shows it is using the very commonly used Nginx web server among web apps.

\footnotesize{\begin{verbatim}
root@ubuntu:~/Desktop/snare$ sudo snare --page-dir 
example.com --tanner 0.0.0.0

   _____ _   _____    ____  ______
  / ___// | / /   |  / __ \/ ____/
  \__ \/  |/ / /| | / /_/ / __/
 ___/ / /|  / ___ |/ _, _/ /___
/____/_/ |_/_/  |_/_/ |_/_____/

    
serving with uuid 593755d8-0aa4-4d88-a970-b5804f4dade7
Debug logs will be stored in /opt/snare/snare.log
Error logs will be stored in /opt/snare/snare.err
======== Running on http://127.0.0.1:8080 ========

\end{verbatim}}

\normalsize
\subsection{\textbf{TANNER: Remote Analysis tool}}
TANNER is the brain of the honeypot, and it analyzes all the requests made through SNARE after that it generates their responses dynamically and sends them back through SNARE. TANNER works asynchronously to increase the overall power and speed of the honeypot. 

\bigskip
\bigskip
\bigskip

\footnotesize{\begin{verbatim}
root@ubuntu:~/Desktop/tanner$ sudo tanner

      _________    _   ___   ____________
     /_  __/   |  / | / / | / / ____/ __ \
      / / / /| | /  |/ /  |/ / __/ / /_/ /
     / / / ___ |/ /|  / /|  / /___/ _, _/
    /_/ /_/  |_/_/ |_/_/ |_/_____/_/ |_|

    
Debug logs will be stored in /opt/tanner/tanner.log
Error logs will be stored in /opt/tanner/tanner.err
======== Running on http://0.0.0.0:8090 ========
(Press CTRL+C to quit)

\end{verbatim}}
\medskip

\normalsize
\subsubsection{\textbf{Database}}

The honeypot uses REDIS as its database for storage, the significant advantages of using key-value storage like Redis are it is incredibly fast, supports multiple data types like hashes, sets, and caching. 
\medskip

\subsubsection{\textbf{Asynchronous Docker}}
One of the advanced concepts in the working of this honeypot is the mechanism to get the injection results from the emulators. Emulators use custom Docker images to get the emulation results of the payload request sent by the attacker. The primary reason behind using Docker for this purpose is the security of the honeypot itself because using the libraries directly from the system could make the honeypot vulnerable, and the chances of the honeypot getting exposed increases.

This mechanism allows the emulators to create and delete the containers automatically and asynchronously for different custom images, behind this mechanism we are using AIODocker (asynchronous Docker) with a separate helper built explicitly for creating and deleting the containers automatically. Aiodocker helper has methods for the creation of containers, fetching the existing container, building the default host image (we can change the default host image using Tanner Config settings), executing the payload of the attacker, and deleting the used container.
\medskip

Different emulators use different images to build the containers, Command Execution emulator uses a busy box docker image, creates the container, executes the attacker payload inside it and then sends the results back to the emulator. Template Injection emulator uses a custom image built with alphine as a base image; this emulator first builds the custom image itself using the Dockerfile and aiodocker helper then installs the supported template engines tornado and mako. This solution makes upgrading the emulator very easy, with the only change needed in the Dockerfile of the custom image to install the new templating engines. At this point, the custom image is responsible for getting the injection results of the tornado and mako templating engines.
\medskip

% Use figure* to make the fig span two columns

\begin{figure*}[htbp]
\centerline{\includegraphics{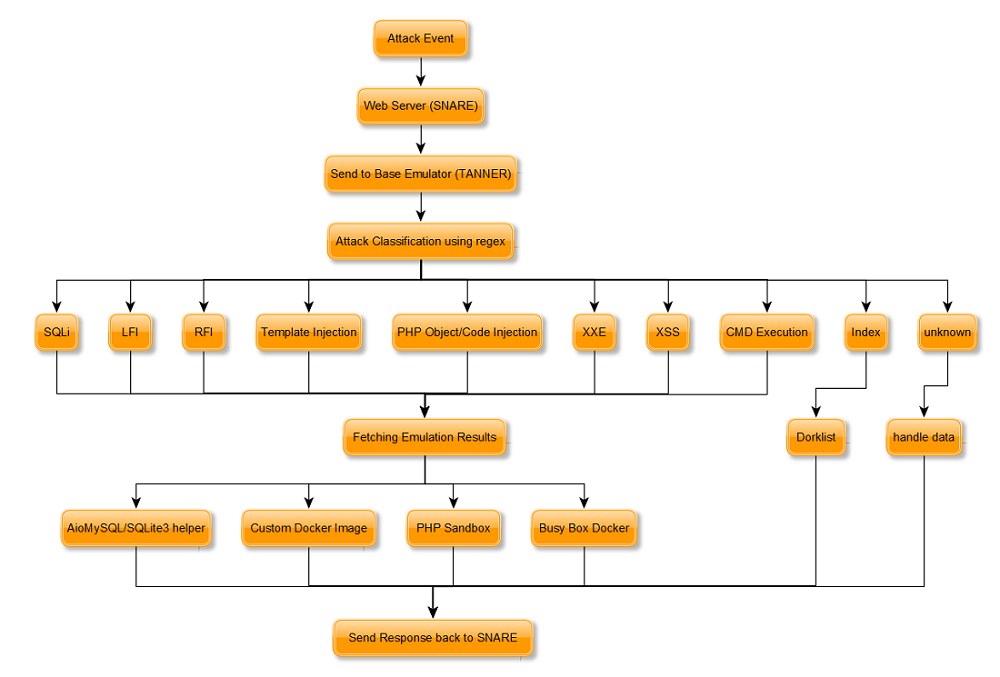}}
\caption{\textit{Flowchart of how an attack gets handled.}}
\label{fig2}
\end{figure*}

\subsubsection{\textbf{Emulation engine}}

These are the heart and soul of the honeypot, every event on SNARE has to be passed through all the emulators these are the attack ships of the honeypot. 

There are three types of emulators - 

\begin{itemize}
\item GET Emulators: These are responsible for analyzing the GET requests. \item POST Emulators: These emulators are responsible for handling the POST requests. They also support GET requests as well.
\item COOKIE Emulators: These emulators support COOKIE attack payloads, GET and POST requests as well. Only the SQLi emulator and PHP Object Injection Emulator support this type.
\end{itemize}

Structure of each emulator has three standard methods - 
\begin{itemize}
\item Scan: This method matches the input payload with the specific regex pattern of the particular vulnerability to check if this attack is legitimate. This method returns a dictionary containing the attack name and attack order.
\item GET Emulation Results: This is the method which gets the emulation results of the payload by various techniques such as-
\begin{itemize}
    \item Injecting the payload into a custom template then sending it to the PHP sandbox (phpox) for the results(applicable for template injection, PHP object/code injection, XXE injection emulators).
    \item Executing the payload directly in a docker container (Command Execution and Local File Inclusion emulators) or by injecting the payload into SQL/MySQL queries and executes them in a dummy database (SQLi and MySQL emulator).
\end{itemize}
\item Handle: This method connects both the above methods and then returns the emulation results in the form of a dictionary containing the results and a boolean parameter-PAGE, which is used to set the injectable page. 
\end{itemize}
\medskip

\subsubsection{\textbf{PHP Sandbox (PHPOX)}}

It is one of the quintessential elements of the honeypot; it is responsible for returning the emulation results for emulators like PHP object/code injection, XXE injection, and Remote File Inclusion (RFI). It is a separate utility from SNARE, and TANNER served separately on a different port. A PHP sandbox helper is built into TANNER to get injection results from the PHPOX. 
\medskip

\subsubsection{\textbf{Base Emulator}}
The mothership of the fleet of attack ship emulator is the BASE Emulator, which manages all the other emulators supporting multiple vulnerabilities.
It uses a set injectable page method to set a custom page in which the final results are injected. Similar to the other emulators Base emulator uses Get Emulation Results method, which first scans the payload through all the emulators to detect the possible vulnerability of the highest order. Then it invokes the handler of the target emulators, which then finds the injection results. This method returns the dictionary containing the attack name, the order of attack, and the injection results. 
\medskip

\subsubsection{\textbf{Template Injection Emulator}}
This emulator imitates the Template Injection vulnerability; it supports tornado and mako templating engines at this point. This emulator first builds the custom host image using a remote path of the Dockerfile (we are using `githubusercontent` remote path for this purpose). The input payload is then matched with the regex pattern to identify the engine type then it is injected into the vulnerable custom template of that engine. The injection formats for the engines are different.
\medskip

For eg: TORNADO: \{\{7*7\}\} renders ``49'' as a result, MAKO: $<$\% x=7*7 \%$>$\$\{x\} also renders ``49''. In the next few releases, we are trying to add support for common engines like PHP's ``twig'' or python's ``jinja2''.

This vulnerable template is executed inside a docker container of the custom image using aiodocker helper, and the handler returns the final results.
\medskip

\subsubsection{\textbf{XML External Entity (XXE) Injection Emulator}}
It emulates XML External Entity Injection vulnerability; it injects the input into a vulnerable PHP template that unloads and parses the XML data from the given document type declaration (DTD). The specialty of this emulator is that it also supports Out of Band(OOB) XXE Injection as well, which can be enabled/disabled from TANNER config settings. The handler then returns the relevant results.
\medskip

\begin{verbatim}
Sample payload: 

<?xml version="1.0" encoding="ISO-8859-1"?>
<!DOCTYPE foo [ <!ELEMENT foo ANY >
<!ENTITY xxe SYSTEM "file:///etc/passwd" >]>
<data>\&xxe;</data>
\end{verbatim}

\medskip

\subsubsection{\textbf{PHP Object/Code Injection Emulator}}
It emulates the PHP Object Injection vulnerability. It occurs during PHP serialization and deserialization of class objects using magic methods like \_sleep, \_destruct, and \_contruct. Attackers directly pass the serialized object based on the already exposed vulnerable code containing magic methods, but the injection results are obtained from PHP sandbox; hence, the attack can be carried out safely like a real system without getting exposed.
\medskip

For PHP code injection methods like eval() are exploited, and it is capable of executing system-level commands.
Vulnerable code - $<$?php eval('\$a = \{payload\}'); ?$>$ The commands in payload are executed which is very dangerous, but similar to object injection this code is executed in PHP sandbox safely.
\medskip

\subsubsection{\textbf{Command Execution Emulator}}
One of the most common injection attacks is Command Injection; this emulator uses a busy box docker image in which the attacker payloads like a cat, echo, cd, ls, ping, and many more run over a bash shell.
\medskip

\subsubsection{\textbf{Attacker/Crawler Detection}}
This feature allows us to detect if the attacker is a crawler or a tool; it uses a binary tree mechanism for the detection process. Every node works as a detection test and assigning a ``cf'' confidence factor after parsing every node. It tries to detect the possible owners: attacker, crawler/tool. 

Attacker detection works by first matching the attack types from LFI, RFI, XXE, and others if resembled then cf=1; otherwise, it checks for session count $>$100 and session duration $<$10, which makes a possibility that the owner is a bot. We use user-agent for matching the bot owners if found then hostname matches for the known bot hosts, success makes cf=0.25 else cf=0.75. If the user-agent can not find the possible hosts, it checks for any hidden links $>$0  which makes cf=0.5.

For detecting crawler/tool, there is ``cf-c'' crawler confidence factor and ``cf-t'' tool confidence factor. It gets the robots.txt file if fetched then cf-c=1 and cf-t=0 as it indicates it is a crawler otherwise it checks for session count $>$100 and session duration $<$10 then it uses user-agent to match bot owners (cf-c=0.85, cf-t=0.15) if not matched use hostname  to match known bot hosts (cf-c=0.75, cf-t=0.15).

\medskip

\section{\textbf{RESULTS AND DISCUSSIONS}}

This section analyses the proposed reactive web application honeypot and its performance in the real world scenario. All the results are obtained by first cloning the domain 'example.com.'  which is reserved by the Internet Engineering Task Force (IETF) for the testing and documentation purposes.

Tanner has a standalone web server for handling and analysing all the incoming requests from snare. First of all, it creates a new session for every attack and displays the statistics. Each session depicts all the vital information about the attack; it tries to detect the attack owner based on the 'confidence factor' and displays as a crawler, an automated tool or a user. Tanner web interface calculates the total number of attack types, confidence factor of user and many more. It also displays other essential parameters which help detect the attack type and attack location, as shown below in Table I.

\begin{table}[htbp]
\caption{Session Information of each attack}
\begin{center}
\begin{tabular}{|c|c|}
\hline
\textbf{Key}&\textbf{Value} \\
\hline
UUID& session-id  \\
\hline
IP address & 192.168.x.x, 127.0.0.1 \\
\hline
Location & attack-directory \\
\hline
Port & 35126, 8080 \\
\hline
User Agents & Mozilla/5.0, Linux x86-64 \\
\hline
Attack Types & lfi, xss, template-injection, sqli, rfi, cmdexec \\
\hline
Possible owners & User:0.15, crawler: 0.5,tool: 0.25 \\
\hline
\end{tabular}
\label{tab1}
\end{center}
\end{table}

All of these statistics are only accessible through the tanner UI for the administrator for monitoring the flow of network requests in real-time. While on the attacker's side, it deceptively acts just like a real system, carrying out the attacks according to the attacker's desires.

The power of tanner's emulation engine helps to generate the responses to the snare events dynamically which in turn increases the stealthiness of the honeypot ten-folds. Behind the scenes, it is using Docker-engine asynchronously injecting the attacker's payloads and getting the results which are as legitimate as a real-world web app attack. Some of the attacks performed during before deploying the honeypot –

\begin{figure}[H]
\centerline{\includegraphics{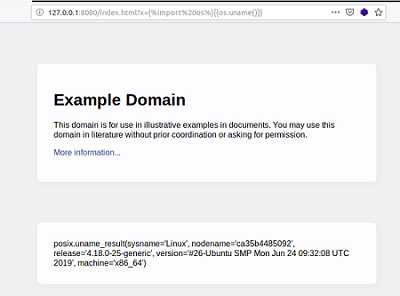}}
\caption{\textit{Template Injection Attack.}}
\label{fig1}
\end{figure}

\begin{figure}[H]
\centerline{\includegraphics{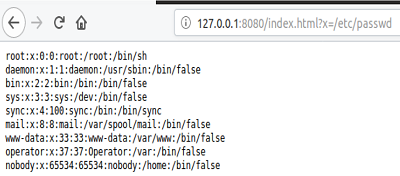}}
\caption{\textit{Local File Inclusion.}}
\label{fig1}
\end{figure}

\section{\textbf{Conclusions and Future Work}}

As compared to other adaptive honeypots like Heat-seeking honeypots, \cite{b11} which fetches the web page from the internet and then puts them on the honeypot. They advertise these web pages on search engines and crawlers to entice attackers, attackers find these indexed pages and try to query them which gets logged in the honeypot. These logs can be used efficiently for further analysis. 

Glastopf, another dynamic and intelligent honeypot, contains web pages inside it, and they are indexed on search engines and crawlers. Our mechanism is a high-level form of Glastopf; it clones the real applications from the internet. It makes constructing web pages less tedious, and it pulls attackers more promptly as it is an exact clone of the real system. It supports more complex real-world vulnerabilities with a robust logging mechanism. 

As a future work of this project, emulation of more complex vulnerabilities, the addition of more templating engines for template injection emulator, and exposing the attacker's identity verbosely.

\vspace{12pt}

\end{document}